\documentclass[aps,pre,a4paper,amsmath,twocolumn,showpacs,floatfix]{revtex4}
\usepackage{graphicx}
\usepackage{amssymb}
\begin{document}

\title{Orientational ordering in hard rectangles: the role of three-body 
correlations} 
\author{Yuri Mart\'{\i}nez-Rat\'on}
\email{yuri@math.uc3m.es}

\affiliation{Grupo Interdisciplinar de Sistemas Complejos (GISC),
Departamento de Matem\'aticas, Escuela Polit\'ecnica Superior,
Universidad Carlos III de Madrid,
Avenida de la Universidad 30, E-28911 Legan\'es, Madrid, Spain.
}

\author{Enrique Velasco}
\email{enrique.velasco@uam.es}

\affiliation{Departamento de F\'{\i}sica Te\'orica de la Materia Condensada
and Instituto de Ciencia de Materiales Nicol\'as Cabrera,
Universidad Aut\'onoma de Madrid, E-28049 Madrid, Spain.}

\author{Luis Mederos}
\email{l.mederos@icmm.csic.es}

\affiliation{Instituto de Ciencia de Materiales, Consejo Superior de
Investigaciones Cient\'{\i}ficas, E-28049 Cantoblanco, Madrid, Spain.}

\date{\today}

\begin{abstract}
We investigate the effect of three-body correlations 
on the phase behavior of hard rectangle two-dimensional fluids. 
The third virial coefficient, $B_3$, is incorporated via an equation 
of state that recovers scaled particle theory for parallel 
hard rectangles. This coefficient, a functional of the 
orientational distribution function, is calculated 
by Monte Carlo integration, using an accurate parameterized distribution 
function, for various particle aspect ratios in the range $1-25$.
A bifurcation analysis of the free energy calculated from the obtained 
equation of state is applied to find the isotropic (I)-uniaxial nematic 
(N$_u$) and isotropic-tetratic nematic (N$_t$) spinodals and to study the 
order of these phase transitions. We find that the relative stability of the 
N$_t$ phase with respect to the isotropic phase is enhanced by the 
introduction of 
$B_3$. Finally, we have calculated the complete phase diagram using a
variational procedure and compared the results with those obtained 
from scaled particle theory and with Monte Carlo simulations carried 
out for hard rectangles with various
aspect ratios. The predictions of our proposed equation of state as
regards the transition densities between the isotropic 
and orientationally ordered
phases for small aspect ratios are in fair agreement with simulations. 
Also, the critical aspect ratio  
below which the N$_t$ phase becomes stable is predicted to increase 
due to three-body correlations, although the corresponding value is
underestimated with respect to simulation.
\end{abstract}

% 64.70.Md  Transitions in liquid crystals
% 64.75.+g  Solubility, segregation, and mixing; phase separation
% 61.20.Gy  Theory and models of liquid structure

\maketitle

\section{Introduction} 

The (two-dimensional) hard-rectangle (HR) model has recently received some
attention due to the possibility that a dense film of these particles
exhibits spontaneous tetratic 
order\cite{Schaklen,Martinez-Raton,Frenkel1,Donev}.
Additional interest originates from
the fact that some types of organic molecular semiconductors are made 
of rectangularly shaped molecules; a notable example is the PTCDA molecule, 
films of which have recently been studied quite intensely \cite{PTCDA}. 
Even though
the interactions between these molecules involve high-order polar forces
(e.g. quadrupolar forces) it is of interest to investigate theoretically
the intrinsic order associated to purely excluded-volume effects with a view 
to predicting structural and thermodynamic properties of the film by
incorporating other interactions via traditional perturbation theories.
The system we investigate in the present work mimics an incommensurate
film of these molecules with only excluded-volume interactions involved
and in the regime where molecules are free to move in the film
(i.e. fluid regime). Phases with two-dimensional crystalline order
will be left for future work.

Recently monolayers of various macroscopically-sized particles have 
been studied using mechanical vibrations on the monolayer to induce
motion \cite{Narayan}. Even though this is an athermal, non-equilibrium 
system reaching steady-state configurations, these configurations
are mainly driven by packing effects and should give the trend as to
what types of order could be expected. In particular, tetratic order was 
observed in particles with sufficiently sharp corners, resembling rectangles,
in contrast with particles such as discorectangles (projections of 
spherocylinders on the plane) which only exhibit nematic ordering,
or basmati-rice grains, which in addition may have a smectic phase.

Apart from the possible interest in modelling the behaviour of monolayers 
made of molecules with technological interest, our investigations have the 
additional, more fundamental aim of 
elucidating the effect that three-body correlations have on the orientational
properties of two-dimensional fluids. Onsager showed that for three-dimensional
hard rod fluids in the limit of infinite aspect ratio (hard-needle limit), 
$\kappa\to\infty$ (with
$\kappa\equiv L/\sigma$, $L$ and $\sigma$ being the length and width of the 
constituent particles), the ratio between the third virial coefficient and 
the second virial coefficient squared asymptotically vanishes, for the
isotropic fluid, as $B_3/B_2^2\sim(\sigma/L)\log(L/\sigma)$ \cite{Onsager}. 
Taking this result into account, he used a second-order virial expansion 
for the free energy as a functional of the orientational distribution function 
and obtained predictions for the isotropic (I)-nematic (N) phase-transition 
densities, exact in the above limit. By contrast, in two dimensions the 
above ratio between virial coefficients has the approximate limiting value of
$0.514$ \cite{Tarjus}, implying that three-body correlations might play a very 
important role in the isotropic fluid even in the hard-needle limit;
this is in sharp contrast with the three-dimensional case. 
An investigation of the
effect of these high-order correlations on orientational ordering seems
therefore appropriate.

Since Onsager theory does not account for higher-than-two body
correlations, an alternative theoretical approach is required.
Scaled particle theory (SPT), first developed for a mixture of hard spheres 
\cite{Reiss} and later extended to anisotropic particles \cite{Cotter,
Lasher,Barboy}, includes as a main ingredient the exact analytic expression 
for the second virial coefficient \cite{Isihara,Kihara}, but again the third 
is approximated assuming $B_3/B^2_2\to 0$ in the hard-needle limit, an
assumption that is incorrect. Also, it has been shown that, for a variety 
of particle shapes in two dimensions, the fourth and fifth virial 
coefficients tend to negative 
values in the same limit \cite{Tarjus,Rigby}. Thus it may very well occur 
that in these cases the virial series exhibits poor convergence, and a
natural question arises: how does the phase behavior of anisotropic 
hard convex two-dimensional fluids change when three- and higher-body 
correlations, neither of which are included in the standard Onsager
and SPT approaches, are taken into account? One of the aims of 
the present article is to shed some light on this question. For this purpose 
we develop an equation of state (EOS) for HR which exactly
includes two- and three-body correlations in the nematic fluid and
recovers SPT in the case of perfectly aligned particles. 

Recent investigations of the HR system have used the SPT approach
\cite{Schaklen,Martinez-Raton} and Monte Carlo (MC) simulations 
\cite{Frenkel1,Donev} to study its phase behavior. Aside from the
usual isotropic-uniaxial nematic (N$_u$) transition, these works
have shown that this peculiar system exhibits a continuous 
transition between the isotropic phase and a tetratic nematic (N$_t$) phase. 
The latter is an orientationally ordered phase but with $D_{4h}$ symmetry, 
i.e., the system is invariant under rotation of $\pi/2$. The SPT predicts 
that this phase is stable up to an aspect ratio $\kappa\approx 2.21$ and 
that the packing fraction values of the I-N$_t$ transition
are around 0.85. This is
in disagreement \cite{Martinez-Raton} with the I-N$_t$ transition densities 
obtained from simulation for $\kappa=1$ and $2$ \cite{Frenkel1,Donev}, which 
predicts values around 0.7.
Considering the importance that high-body correlations may have 
on the phase behavior of two-dimensional hard-convex bodies, it is
the second purpose of this article to apply our model (which 
includes three-body correlations) to calculate the phase diagram of HR 
and compare the results with those of SPT and MC simulations. The main features
of the phase diagram are calculated using bifurcation theory for the I-N 
transition and also minimizing the nematic free energy functional 
resulting from our proposed EOS.

The article is organized as follows. In Section II we describe the  
theoretical model for a general two-dimensional hard-convex fluid
as applied to the isotropic and nematic fluids. Section III is devoted to 
the results obtained from the analysis of the theory, and comparison
is made with simulation results; also, the complete liquid-crystal
phase diagram is presented. Finally, some conclusions 
are drawn in Section IV. The Appendix contains a detailed account of
the bifurcation analysis and the minimization method used to analyse
the phase behaviour of the model, together with some details on the
computer simulations.

\section{Theory}
In the present section we introduce the theoretical formalism necessary
for the study of the phase behavior of the HR fluid. This includes the
derivation of the EOS for the isotropic and nematic fluids, 
along with the corresponding free energy density. The phase behaviour of the 
model, to be presented in the next section, is analysed by means of two 
complementary techniques: a bifurcation analysis of the free energy density, 
and a full minimization using an accurate functional form for the orientational
distribution function. Details on these techniques are given in the Appendix.

\subsection{EOS for the isotropic fluid}

In this section an equation of state for the isotropic phase, to be extended
later to the nematic phase, is proposed, on a somewhat ad-hoc, but at the 
same well-founded, basis. The virial coefficients are intimately related
to geometric properties of the planar objects making up the fluid.
The second virial coefficient of planar hard particle has the analytic 
form \cite{Kihara}
\begin{eqnarray}
B_2=v+\frac{
{\cal L}^2}{4\pi},
\label{segundo}
\end{eqnarray}
where $v$ and ${\cal L}$ are the area and perimeter of the particle. 
Some approximate analytic expressions for third virial coefficients 
of isotropic fluids made of three-dimensional bodies, as a function of 
their volume, surface area and mean curvature, have been proposed 
\cite{Boublik}. When compared with results from numerical calculations 
\cite{Boublik}, some of these expressions are seen to constitute accurate 
approximations. In two dimensions volume has to
be substituted by area, area by perimeter and mean curvature by a function 
proportional to the perimeter (the latter is true for the most representative 
two-dimensional convex bodies, i.e. rectangles, discorectangles, 
and ellipses). Following some of the most successful approximations 
\cite{Boublik} but translated to the two-dimensional case, we write the 
following analytic expression for the third virial coefficient        
\begin{eqnarray}
B_3=v^2+
\frac{\alpha}{4\pi}v{\cal L}^2+
\frac{\beta}{(4\pi)^2}{\cal L}^4
\label{tercero}
\end{eqnarray}
where the numerical coefficients $\alpha$ and $\beta$ are chosen in 
such a way as to guarantee i) the correct asymptotic hard-needle limit, and 
ii) a good comparison with well-known EOS for some isotropic fluids,
such as hard disks. For the latter we have ${\cal L}^2=4\pi v$ so that
the three terms in the right-hand side of Eqn. (\ref{tercero}) can be 
unified into the single term $(1+\alpha+\beta)v^2$.
The SPT for hard disks is recovered by choosing $\alpha+\beta=2$, whereas
the SPT form for $B_3$ for a general anisotropic particle 
is obtained from (\ref{tercero}) with $\alpha=2$ and $\beta=0$. 
Note, from Eqns. (\ref{segundo}) and (\ref{tercero}), that
\begin{eqnarray}
B_2\sim \frac{{\cal L}^2}{4\pi},\quad 
B_3\sim\frac{\beta}{(4\pi)^2}{\cal L}^4,\label{undos}
\end{eqnarray}
in the infinite aspect-ratio limit, 
as the particle area is proportional to the product 
of the two characteristic lengths of the particle 
(the width $\sigma$ and the length $L$, the latter being 
the larger one), while the perimeter is proportional to their sum. 
Therefore we obtain the asymptotic limit 
$B_3/B_2^2\to \beta$ when $L/\sigma\to\infty$ and
a sensible choice is $\beta=0.514$, the exact asymptotic value of
this ratio \cite{Tarjus}. 

The EOS is obtained by imposing two requirements:
(i) the divergence of pressure at 
high packing fractions is of the form $\sim (1-\eta)^{-2}$ as stated by SPT, 
and (ii) the second and third virial coefficients are obtained from the 
exact virial expansion  
\begin{eqnarray}
\beta P&=&\rho+\rho^2B_2+\rho^3B_3,\label{virial}
\label{lala}
\end{eqnarray}
(where $\rho$ is the density of particles). In other words, we require that 
the third-order virial expansion of the interaction part of the EOS,
\begin{eqnarray}
\beta P_{\rm{exc}}v=\beta P v-\eta=\frac{a_2\eta^2+a_3 \eta^3}{(1-\eta)^2},
\end{eqnarray}
($\eta=\rho v$ being the packing fraction) 
coincides with the exact one (\ref{lala}).
This allows us to obtain $a_k$ ($k=1,2$) as $a_2=1+b_2$, and $a_3=b_3-2a_2-1$, 
where the new coefficients
\begin{eqnarray}
b_k=\frac{B_k}{v^{k-1}}-1,\quad k=2,3,
\end{eqnarray}have been defined in terms of the virial coefficients $B_k$. 
The resulting EOS has the form 
\begin{eqnarray}
\beta Pv=\frac{\eta}{1-\eta}+\frac{\eta^2}{(1-\eta)^2}+
\frac{\eta^3}{(1-\eta)^2}\left(b_3-2b_2\right)
\label{note}
\end{eqnarray}
From this EOS the free energy density can be obtained: we first write
\begin{eqnarray}
\rho^2\frac{\partial \varphi}{\partial\rho}=\beta P(\rho),
\label{porco}
\end{eqnarray}
where $\varphi=\varphi_{\rm{id}}+\varphi_{\rm{ex}}$ is the free energy per
particle and $\varphi_{\rm{id}}$ and $\varphi_{\rm{ex}}$ the corresponding 
ideal and excess contributions. Now using $\beta P$ from (\ref{note}), 
Eqn. (\ref{porco}) can be integrated to give 
\begin{eqnarray}
\varphi_{\rm{exc}}&=&-\ln(1-\eta)+\frac{\eta}{1-\eta} b_2+
(b_3-2b_2)\theta(\eta),\label{taken}\\
\theta(\eta)&=&\frac{\eta}{1-\eta}+\ln(1-\eta)
\label{exceso}
\end{eqnarray}

The first two terms of (\ref{note}) and (\ref{taken}) are SPT-like terms. 
Eq. (\ref{note}), with the exact second and third virial 
coefficients for the particular case of parallel hard rectagles, recovers 
the SPT result [this is easily obtained if we substitute the exact values 
$B_2=2v$ ($b_2=1$) and $B_3=3v^2$ ($b_3=2$) in Eq. (\ref{note})]. 

Inserting $B_2$ and the approximation for $B_3$ from 
(\ref{segundo}) and (\ref{tercero}), respectively, 
in Eqns. (\ref{note}) and (\ref{taken}), 
we obtain our proposed EOS and the excess part of the free energy 
per particle for the isotropic fluid as 
\begin{eqnarray}
\beta Pv&=&\frac{\eta}{1-\eta}+\frac{\eta^2}{(1-\eta)^2}
\gamma\left[1+(\alpha-2+\beta\gamma)\eta\right],\label{state}\\
\varphi_{\rm{ex}}&=&-\ln(1-\eta)+\frac{\gamma\eta}{1-\eta}
+\gamma(\alpha-2+\beta\gamma)\theta(\eta),\nonumber\\ 
\end{eqnarray} 
where 
the anisometric parameter $\gamma={\cal L}^2/(4\pi v)$ was
defined. Note that 
for $\alpha=2$, $\beta=0$, this equation recovers the SPT expression
for hard convex bodies.
From (\ref{state}) 
the following expression for the reduced virial coefficients is obtained:
\begin{eqnarray}
B_n^*\equiv\frac{B_n}{B_2^{n-1}}=\frac{1+\left[1+(\alpha-1)(n-2)\right]
\gamma+\beta(n-2)\gamma^2}{(1+\gamma)^{n-1}}\nonumber\\
\label{bast}
\end{eqnarray} 

\subsection{EOS for the nematic fluid}

The EOS for the nematic fluid is now obtained from 
Eqn. (\ref{note}) by substituting the virial coefficients of the
isotropic fluid $B_{n}$ ($n=2,3$) by their functional versions 
$B_{n}[h]$ for the nematic fluid; here $h(\phi)$ is the orientational
distribution function.
The latter coefficients are obtained from the definitions of $B_2$ and $B_3$, 
in terms of integrals over the Mayer function:
\begin{eqnarray}
&&B_k[h]=\frac{1}{k}\left[\prod_{l=1}^k\int d\phi_lh(\phi_l)\right]
{\cal K}(\phi_1,\dots,\phi_k),\\
&&{\cal K}(\phi_1,\phi_2)=-\int d{\bf r}f({\bf r},\phi_{12}),\\
&&{\cal K}(\phi_1,\phi_2,\phi_3)=-
\int d{\bf r}\int d{\bf r}'f({\bf r},\phi_{12})f({\bf r}',\phi_{23})
\nonumber \\&&\times f({\bf r}-{\bf r}',\phi_{13}),\label{kernel3}
\end{eqnarray}
where $\phi_{\alpha\beta}=\phi_{\alpha}-\phi_{\beta}$ is the relative 
angle between axes of particles $\alpha$ and $\beta$, and 
$f({\bf r},\phi_{\alpha\beta})$ the Mayer function. The corresponding
free-energy functional $\varphi[h]=\varphi_{\rm{id}}[h]+
\varphi_{\rm{ex}}[h]$ is obtained from Eqns. (\ref{taken}):
\begin{eqnarray}
\varphi_{\rm{exc}}[h]=-\ln(1-\eta)+\frac{\eta}{1-\eta} b_2[h]+ 
\left(b_3[h]-2b_2[h]\right)\theta(\eta)\nonumber\\
\label{taken1}
\end{eqnarray}
with the ideal part exactly calculated from 
\begin{eqnarray}
\varphi_{\rm{id}}[h]=\ln \eta -1 +\int_0^{2\pi}d\phi h(\phi) \ln\left[
2\pi h(\phi)\right].
\label{id}
\end{eqnarray}
 The remaining virial  
coefficients are approximated from Eq. (\ref{note}) by 
\begin{eqnarray}
B_n[h]=v^{n-3}\left\{(n-2)B_3[h]-(n-3)B_2[h]v\right\}.
\end{eqnarray}  

The integral over spatial variables in the definition of $B_2[h]$ 
is known analytically for most convex bodies. In particular, for HR we have 
\begin{eqnarray}
&&{\cal K}(\phi_1,\phi_2)=
\left(L^2+\sigma^2\right)|\sin \phi_{12}|+
2L\sigma \left(1+|\cos\phi_{12}|\right),\nonumber \\
\end{eqnarray}
which is the excluded area between particles with relative orientation 
$\phi_{12}$. However, the required double angular average over $h(\phi)$
has to be estimated numerically (we used Gaussian quadrature). Also, in 
the case of $B_3[h]$, all integrals have to be calculated numerically
(using MC integration). For this purpose we found
it convenient to use a parameterized orientational distribution function
\begin{eqnarray}
h(\phi)=C \exp{\left(\sum_{\tau=1}^n\lambda_{\tau}
\cos(2\tau\phi), 
\right)}
\end{eqnarray}
in terms of the $n$ parameters $\lambda_{\tau}$ ($\tau=1,\dots,n$). 
$C$ is a normalization constant.
In practice two variational parameters ($n=2$) were used. 
Details on how this calculation was realized in practice are relegated to 
the Appendix. 

\section{Results}
\begin{figure}
\mbox{\includegraphics*[width=2.9in, angle=0]{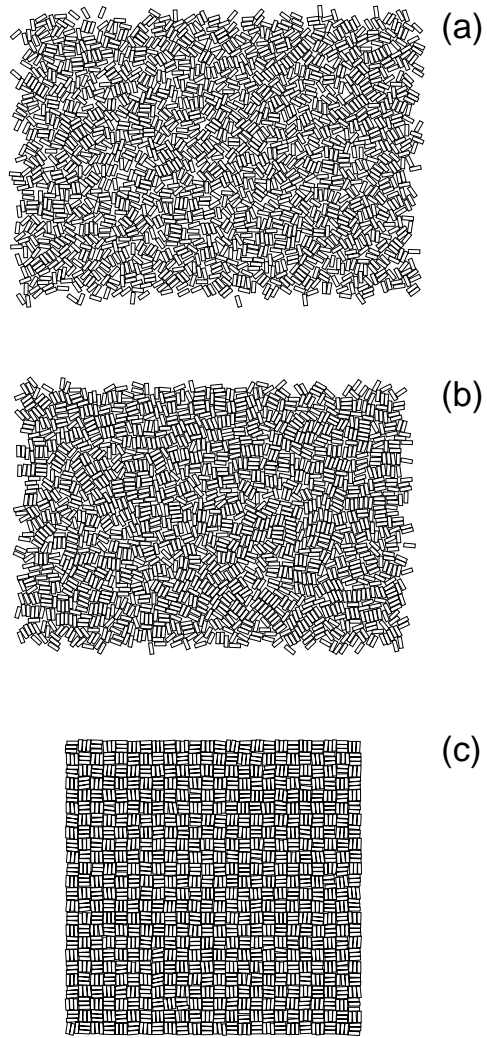}}
\caption{Typical particle configurations as obtained from MC simulations.
(a) isotropic phase; (b) tetratic phase; and (c) crystalline phase
with particles arranged in one of the possible configurations. See
text for details on the simulations.}
\label{config}
\end{figure}

In this section we present the main results obtained from the inclusion 
of three-body correlations into the EOS for the isotropic and the nematic 
fluids, as proposed in Section II. The results from bifurcation analysis 
and from numerical minimization of the free-energy functional are presented 
in Section III B. In the latter case we compare the results from the present 
theory with those obtained from SPT and from simulations. But before 
presenting the results, we show in Fig. \ref{config} a series of
particle snapshots extracted from our simulation runs. Details on the
simulations are given later on. The three configurations are representative
of an isotropic phase [Fig. \ref{config}(a)], a tetratic phase 
[Fig. \ref{config}(b)], and a crystalline phase [Fig. \ref{config}(c)].

\subsection{Isotropic fluid}

\begin{figure}
\mbox{\includegraphics*[width=3.4in, angle=0]{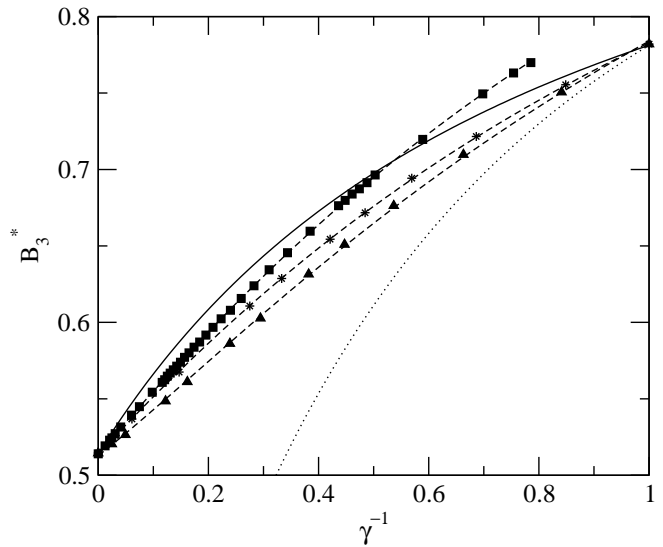}}
\caption{Reduced virial coefficient $B_3^*$ 
as a function of the anisometric parameter $\gamma$. 
Simulation results are shown for 
hard rectangles (squares), 
discorectangles (asterisk), and ellipses (triangles). The solid and dotted 
lines are the results from Eq. (\ref{bast}) 
with $(\alpha,\beta)=(1.611,0.514)$, 
and $(\alpha,\beta)=(2,1/8)$ respectively. Also are shown with dashed lines 
the best fits from Eq. (\ref{misma}).}    
\label{fig1}
\end{figure}

Let us first compare the different approximations for $B_3$ and assess
their quality according to the degree of agreement with MC-integration
results for isotropic fluids made of different hard-convex bodies.
Our own MC calculations have been carried out for hard rectangles, while those
of Ref. \cite{Rigby} were focussed on hard discorectangles and hard ellipses.  
All the results are plotted in Fig. \ref{fig1} along with
two different analytic approximations for the reduced virial coefficient 
$B_3^*$. One of them (solid line) is calculated from Eqn. (\ref{bast}) with 
$\beta=0.514$ [which gives the correct value of $B_3$ in the Onsager limit 
\cite{Rigby} --see Eqn. (\ref{undos})], and setting $\alpha=1.611$, which 
gives the correct third virial coefficient for hard disks 
($B_3^*\approx3.125$). The other (dotted) line is also calculated from
Eqn. (\ref{bast}), but choosing $\alpha=2$ and $\beta=1/8$, 
which reduces to the proposal made by Boublik in Ref. \cite{Boublik1}. 
As will be shown below, this proposal approximates the EOS for the
isotropic phase of hard convex bodies reasonably well.
Also plotted in Fig. \ref{fig1} with dashed lines are the best fits 
calculated from 
\begin{eqnarray}
B_3=\delta_{\rm{HB}} v^2+\frac{\alpha_{\rm{HB}}}{4\pi}v{\cal L}^2+
\frac{\beta_{\rm{HB}}}{(4\pi)^2}
{\cal L}^4
\label{misma}
\end{eqnarray}
with values for the coefficients $\delta_{\rm{HB}}$, $\alpha_{\rm{HB}}$, 
and $\beta_{\rm{HB}}$ depending 
on the particle geometry  ($\rm{HB}\equiv HR,HDR,HE$), i.e. hard rectangles, 
hard discorectangles, and hard ellipses. Note that Eqn. (\ref{misma}) is 
the same as Eqn. (\ref{tercero}), but with a new numerical coefficient 
$\delta_{\rm{HB}}$ as a prefactor of $v^2$. From Fig. \ref{fig1} 
we can see that the present approximation ($\alpha=1.611,\beta=0.514$) 
describes the behavior of $B_3$ as a function of the anisometric parameter 
$\gamma$ much better than Boublik's proposal which, in the Onsager limit,
gives the (wrong) value $B_3^*=1/8$. Also, if one is to describe the correct 
behavior of $B_3^*$ for different particle geometries in the whole range 
of $\gamma$, it is necessary to take $\delta_{\rm{HB}}\neq 1$.

\begin{figure}
\mbox{\includegraphics*[width=3.4in, angle=0]{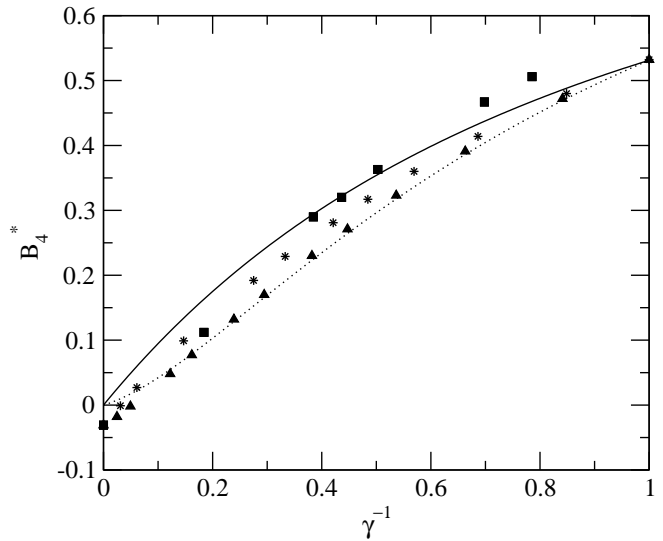}}
\caption{Reduced virial coefficient $B_4^*$ 
as a function of the anisometric parameter $\gamma$. 
All the lines and symbols label the same as in Fig. \ref{fig1}.} 
\label{fig2}
\end{figure}

Numerical values for the coefficients $B_4^*$ and $B_5^*$ have been
calculated in Ref. \cite{Rigby}, for three different particle geometries,
using MC integration. The results for $B_4^*$ are shown in Fig. \ref{fig2}. 
Also plotted are our analytic proposal (solid line) and that of Boublik 
(dotted line). For HR with small anisometry values our approximation is
better than that of Boublik, while the opposite occurs for high 
anisometries. For hard ellipses, Boublik's approach describes reasonably
well the behavior of $B_4^*$ in the whole range of $\gamma$ (except for 
very long particles which have negative values of $B_4^*$). 

\begin{figure}
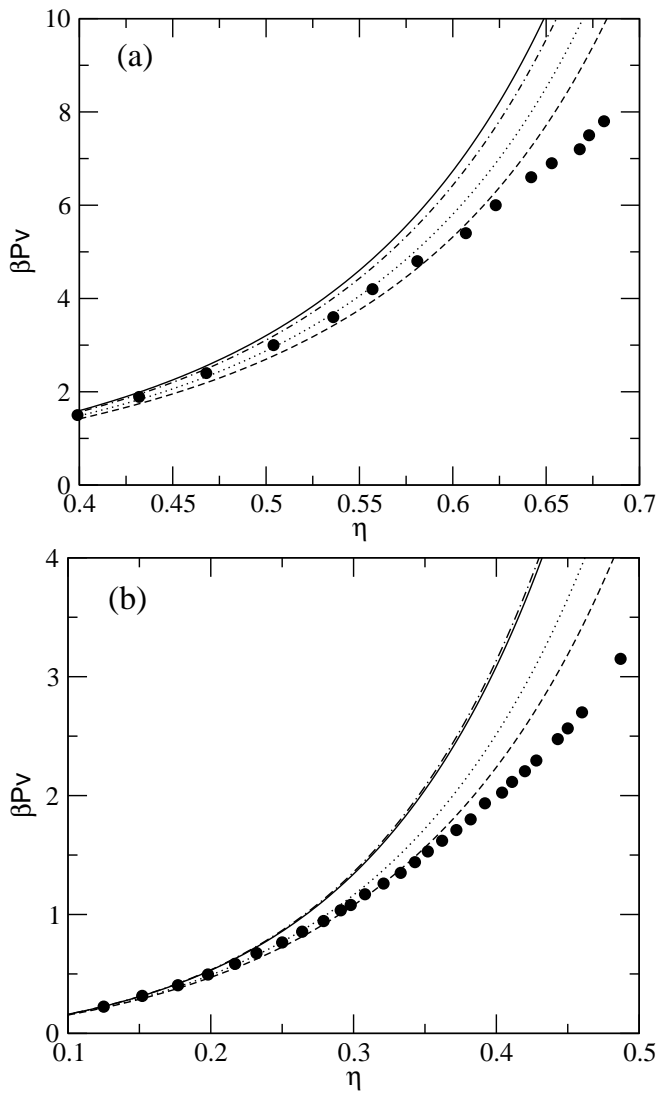

\mbox{\includegraphics*[width=3.4in, angle=0]{Fig4.eps}}
\mbox{\includegraphics*[width=3.4in, angle=0]{Fig5.eps}}
\caption{Results from MC simulations (filled circles) on a system of 
10800 HRs with $\kappa=3$ (a), and $\kappa=9$ (b). Dot-dashed line: 
Eq. (\ref{note}) with $B_3$ calculated from MC integration, dashed line: 
SPT, dotted line: Boublik proposal, and solid line: our proposal 
[The $B3$ from Eq. (\ref{tercero}) with $\{\alpha=1.611,\beta=0.514\}$].}   
\label{fig3}
\end{figure}

The EOS obtained from the above approximations can be checked against
MC simulations of systems of HR particles. In order to realize this
comparison we have carried out constant-pressure MC simulations on 
systems of HR with different aspect ratios in the range of pressures 
where the isotropic fluid is the stable phase. The results for $\kappa=3$ 
and 9 are shown in Figs. \ref{fig3} (a) and (b), respectively. Simulations 
were done on systems of $\sim 10^3$ particles, equilibrated along typically 
$\sim 10^6$ MC steps, and averaging over $\sim 4\times 10^6$ MC steps. 
The system was prepared in each case in a crystalline 
low-density configuration with
perfectly aligned particles at low pressure. This configuration rapidly 
turned into a disordered configuration, which was then equilibrated.
After averaging, the system was subject to a higher pressure and then
equilibrated, and the process was repeated increasing the pressure. 
In this way the EOS in the entire region of isotropic stability was obtained.
For comparison we also show in Figs. \ref{fig3} (a) and (b)
the EOS corresponding to 
the SPT (dashed line), Boublik proposal (dotted line), 
our proposal (solid line) and the EOS [Eqn. (\ref{note})] with the virial 
coefficient $B_3$ calculated from MC integration.  
As can be seen the SPT and Boublik's proposal approximate better the 
simulation results. However, given that both theories make wrong predictions 
of the behavior of the third and fourth virial coefficients of HR as a 
function of the anisometric parameter, these results are to be taken with 
caution in the sense that they could be a mere coincidence. We can also see 
from the figures that 
our proposal overestimates the pressure. This kind of behavior is typical 
of fluids composed of hard-core particles which exhibit poorly convergent 
virial series; this seems to be the case for HR particles since
their fourth and fifth virial coefficients become negative for high 
anisometries.

\subsection{Bifurcation to nematic fluid}

We implemented numerically the bifurcation-theory analysis, described 
in detail in the Appendix, to calculate the spinodal instabilities from the 
isotropic phase to the N$_u$ and N$_t$ phases, and elucidated 
the order of these transitions within the same formalism. These results
were checked against a full minimisation of the free-energy functional
employing the methodology outlined in Section IIB, which in addition
enabled the N$_t-$N$_u$ spinodals, which cannot be easily calculated 
using bifurcation theory, to be obtained. Also, in order to have
essentially exact results for the phase behaviour of this system,
we performed constant-pressure MC simulations on systems of $\sim 10^3-10^4$ 
HR particles and obtained the equations of state and orientational order
(details on these simulations are included in Section D of the Appendix).
All of these results are described in the following.

The I-N$_u$ and I-N$_t$ spinodal lines
$\eta^*(\kappa)$ (the packing fraction at bifurcation as a function
of the aspect ratio $\kappa$) were calculated by solving Eqn. (39)
for $y=\eta/(1-\eta)$ (or $\eta$) for a discrete set of values of $\kappa$
(see Appendix). All the coefficients $b_3^{(k1,k2,k3)}$ that enter
this equation were calculated via MC integration; typically
$\sim 10^8$ MC steps were used to evaluate these coefficients. To elucidate
the order of transitions, we solved Eqn. (46) to find (i)
the value $\kappa_1$ at which the free energy difference
between N$_u$ or N$_t$ and isotropic phases
changes from negative to positive, which in turn
reflects the change of sign of
the coefficient $B^*$ (see Appendix), and (ii) the value $\kappa_2$ for
which the inverse of the isothermal compressibility of the $N_u$ or
$N_t$ phases [$\left(\kappa^{-1}_{\rm{N}}v\right)^*$] at the
bifurcation point becomes zero.
Again, all the coefficients
$b_3$ (the rescaled third virial coefficient) and $b_3^{(k1,k2,k3)}$,
necessary to solve Eqn. (46), were evaluated using MC integration with
the same number of steps as previously. The
quantities $B^*$ and $\left(\kappa^{-1}_{\rm{N}}v\right)^*$ are
shown as a function of $\kappa$ in Figs. 4 (a) and (b) for the
I-N$_u$ transition, and in Figs. 5 (a) and (b) for the I-N$_t$
transition.

\begin{figure}
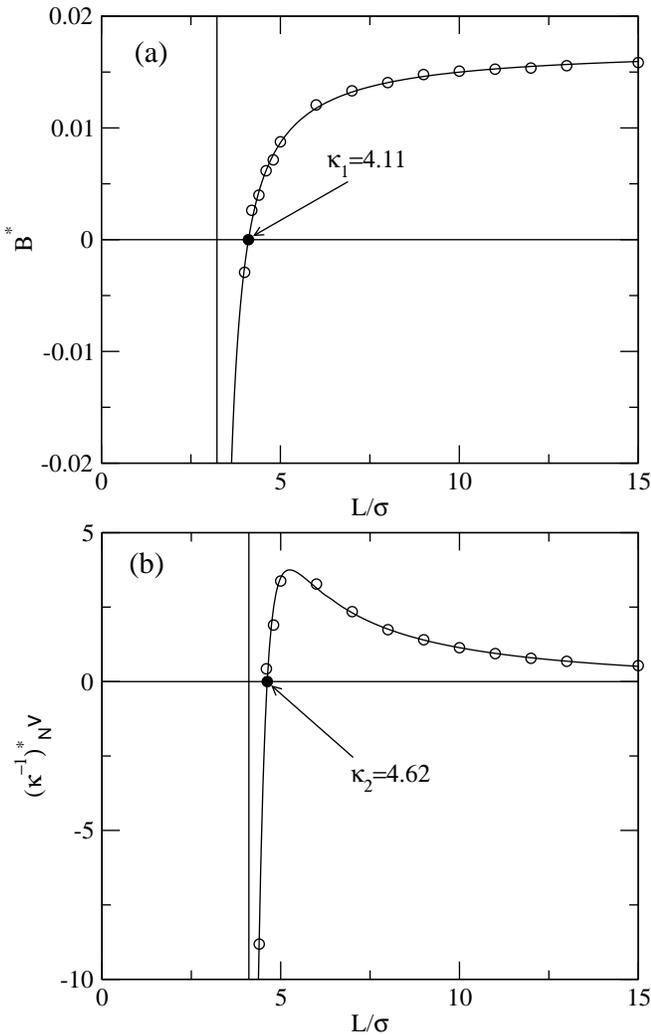

\mbox{\includegraphics*[width=3.4in, angle=0]{Fig6.eps}}
\hspace*{0.02cm}
\mbox{\includegraphics*[width=3.35in, angle=0]{Fig7.eps}}
\caption{The coefficient $B^*$ (a), and the inverse of the 
compressibility factor $\left(\varkappa_{\rm{N}}^{-1}v\right)^*$ (b) at the 
I-N$_u$ bifurcation point as a function of the aspect ratio calculated 
for a discrete set of values (open circles). The filled circles indicate 
the value of $\kappa$ for which they become zero. Thus, 
$\kappa^*=\kappa_2\approx 4.62$ is the true tricritical point.}
\label{fig4}
\end{figure}

\begin{figure}
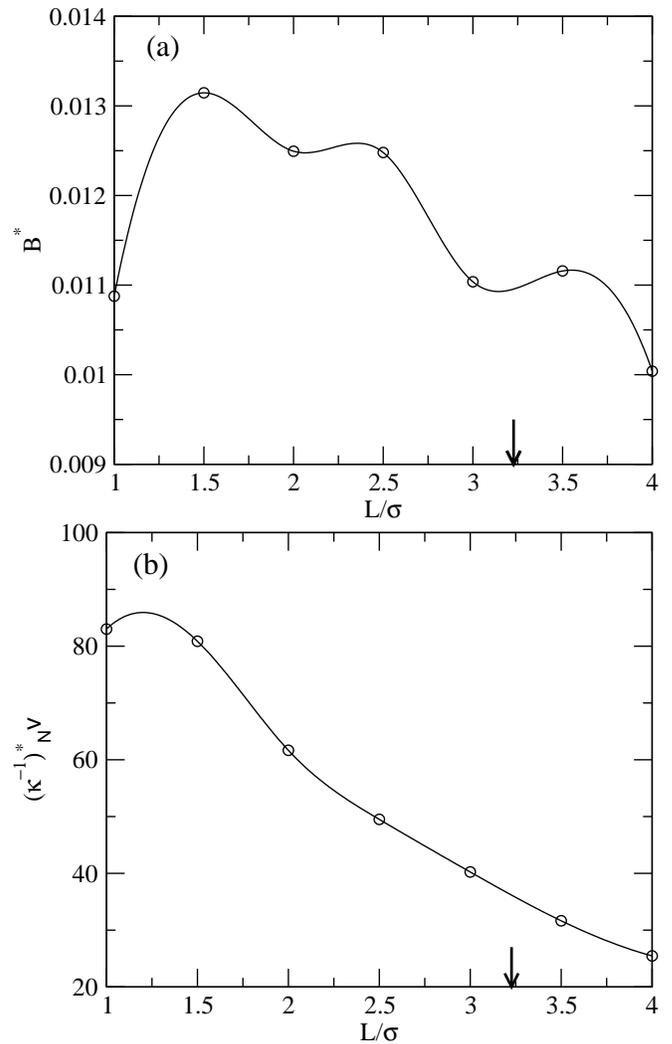

\mbox{\includegraphics*[width=3.4in, angle=0]{Fig8.eps}}
\mbox{\includegraphics*[width=3.4in, angle=0]{Fig9.eps}}
\caption{The coefficient $B^*$ (a), and the inverse of the 
compressibility factor $\left(\varkappa_{\rm{N}}^{-1}v\right)^*$ (b) at the 
I-N$_t$ bifurcation point as a function of the aspect ratio calculated 
for a discrete set of values (open circles). Both magnitudes are always 
greater than zero, so the I-N$_t$ is always of second order. The arrow 
indicates the maximum aspect ratio of N$_t$ phase stability.}  
\label{fig5}
\end{figure}

As can be seen from the figures, $\left(\varkappa_{\rm{N}}^{-1}v\right)^*$
first changes sign from positive to negative at $\kappa_2\approx 4.62$, and 
then diverges at $\kappa\approx 4.11$, coinciding with $\kappa_1$, the 
zero of $B^*$ [see Fig. \ref{fig4} (a)]. The latter has a pole at $\kappa=3.23$,
which is the intersection point between the I-N$_u$ and I-N$_t$ spinodals (for 
larger values of $\kappa$ the I-N$_u$ spinodal lies below the I-N$_t$ spinodal).
This result can be understood from the definition of $B^*$ 
[see Eq. (\ref{otrab})], which diverges at $D^*=0$; this in turn coincides 
with the condition $A^*=0$ if we change $k$ to $2k$ in Eqn. (\ref{laa}) 
to include tetratic symmetry. Thus $B^*$ as a function of $\kappa$ should 
diverge at the point where the I-N$_u$ and I-N$_t$ spinodals intersect. 
The values of $B^*$ and $\left(\varkappa_{\rm{N}}^{-1}v\right)^*$ 
as a function of $\kappa$, calculated this time at the I-N$_t$ spinodal, 
are shown in Fig. \ref{fig5} (a) and (b). From the figure we can see 
that they are always positive, in particular for values of $\kappa$ less 
than $3.23$. Thus we can conclude that the I-N$_t$ transition is always of 
second order. We also note the oscillatory behavior of
$B^*$ as a function of $\kappa$ [see Fig. \ref{fig5}(a)]; this feature
has been shown not to be a consequence of numerical errors inherent to
our MC integration: MC estimates of the coefficients involved in the definition 
of $B^*$ were obtained by increasing the number of MC steps from $10^6$ to 
$10^9$, and the oscillating behavior remained.

\subsection{Phase diagram}

\begin{figure}
\mbox{\includegraphics*[width=3.4in, angle=0]{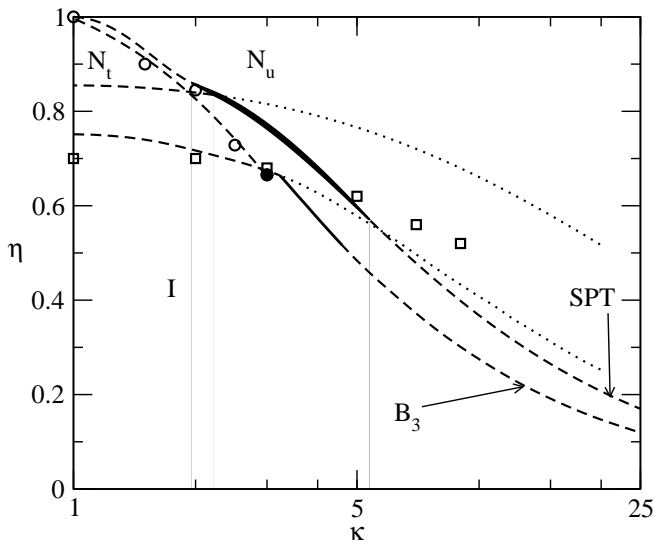}}
\caption{Phase diagram of the HR fluid. 
Continuous and first-order transitions are indicated by
dashed and solid lines, respectively. Dotted lines indicate extension 
of I-N$_t$ line into region where N$_t$ is preempted by uniaxial nematic
phase. SPT transition lines and $B_3$ spinodals are indicated by 
corresponding labels. Circles: minimisation of free-energy functional 
in $B_3$ theory, giving first-order (filled circles) or second-order
(open circles) transitions. Open squares: simulation results for the 
isotropic-to-nematic transition.}
\label{diagramafase}
\end{figure}

The resulting spinodal instabilities from the I to the orientationally 
ordered phases, as calculated from the bifurcation analysis,
are shown in Fig. \ref{diagramafase}. Also shown in the same figure is the
complete phase diagram resulting from SPT, already calculated 
in Ref. \cite{Martinez-Raton}. As can be seen from the figure, the inclusion 
of three-body correlations considerably lowers the transition 
densities between isotropic and the orientationally ordered phases. The new   
results compare fairly well with those from MC simulations in the region of
low particle aspect ratio (our simulations for 
$\kappa=$3, and simulations for $\kappa=1$ 
\cite{Frenkel1} and $\kappa=2$ \cite{Donev}), represented in the figure
by open squares. Another interesting 
point to remark is that the critical value of $\kappa$ 
below which the N$_t$ phase is stable increases from $\kappa=2.62$ in SPT to 
$\kappa=3.23$ in the new theory. Finally, the I-N$_u$ 
tricritical point occurs at $\kappa^*=\rm{max}(\kappa_1,\kappa_2)\approx 
4.62$ [see Fig. \ref{fig5} (a) and (b)], 
which is lower than the SPT result ($\kappa^*=5.44$). Thus, we can conclude that 
three-body correlations have the effect of lowering the transition densities, 
increasing the stability of the N$_t$ phase and making the I-N$_u$ transition
weaker. The enhanced stability of the tetratic phase is in agreement with simulation 
results.

\begin{figure}
\mbox{\includegraphics*[width=3.4in, angle=0]{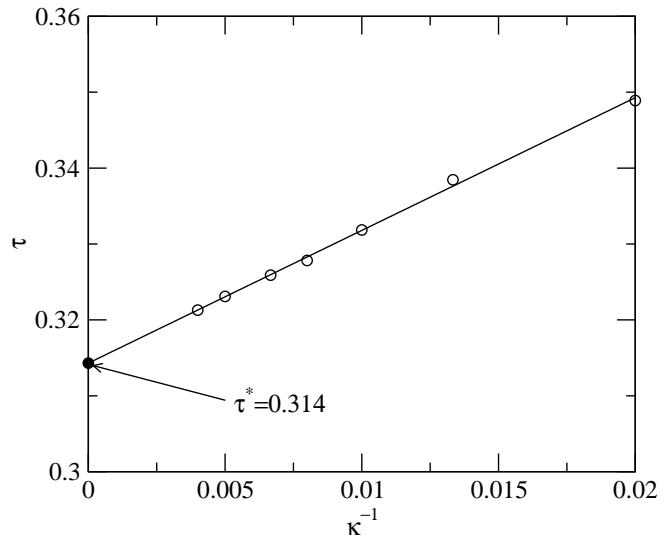}}
\caption{The coefficient $\tau(\kappa)$ [see Eq. (\ref{tau})] as a function of 
$\kappa$ for a discrete set of $\kappa$'s (open circles) calculated 
via Monte-Carlo integration. The straight line calculated from 
mean square approximation intersects the ordinate at the value indicated 
in the figure.}
\label{fig6a}
\end{figure}

It is also apparent from Fig. \ref{diagramafase} that the I-N$_u$ transition 
predicted by the new theory for high values of $\kappa$ occurs 
at packing fractions below those
predicted by SPT. In the Onsager limit, the reduced transition density 
$\rho_r=\rho^* L^2$ for the I-N$_u$ transition can be calculated 
within the third virial-coefficient approximation [see Eqn. (\ref{rho_r}) of
the Appendix].
This reduced density depends on the coefficient $\tau^*$ defined in Eqn. 
(\ref{tau}), which can be calculated by extrapolating the data for 
$\tau(\kappa)$ obtained from simulations. These data, shown in Fig. 
\ref{fig6a} for high values of $\kappa$, are fitted very accurately by
means of a straight line that intersects the vertical axis at 
$\tau^*\approx 0.314$. 
Inserting this value in (\ref{rho_r}), we obtain $\rho_r^{\rm{B_3}}=3.15$, 
which is less than the SPT result $\rho_r^{\rm{SPT}}=4.71$ and much less 
than the MC simulation value, which has been estimated to be between 
7 and 7.5 \cite{Frenkel1}. 
This disagreement is probably due to the poorly convergent character of the 
virial series. As already pointed out, the fourth and fifth virial coefficients
are negative in this limit, so the proper inclusion of higher-order virial 
coefficients is necessary in order to obtain an accurate approximation for the 
I-N transition densities. 

Another interesting aspect of the phase diagram is the failure of the new theory
to reproduce the transition from the isotropic to the nematic phase in the
range of large aspect ratios explored by our simulations. 
Note that, as will be discussed later, the simulations 
cannot reach any definite conclusion as to the real nature (whether tetratic or uniaxial) 
of the nematic phase, especially for large aspect ratio. The fact
that the isotropic-to-nematic transition line $\eta(\kappa)$ obtained from simulations 
in the range $\kappa=1-9$ is a smoothly decreasing monotonic curve and that this
line is quite close to the spinodal line for the I-N$_t$ transition obtained from
the new theory in the same range of aspect ratios, may be indicating that the 
stability of the uniaxial phase is largely overestimated by the new theory, but that
tetratic ordering is relatively well reproduced. This is simply a hypothesis not
based on any real evidence.

\subsection{Further results}

\begin{figure}
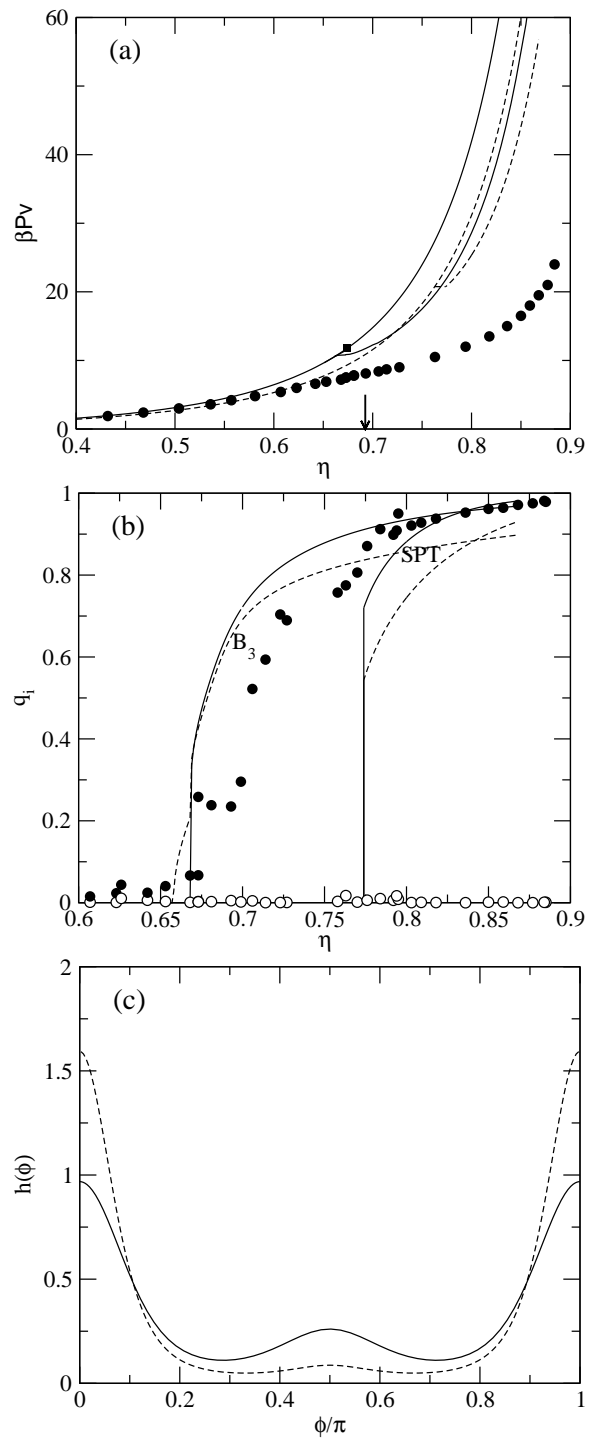

\mbox{\includegraphics*[width=3.in, angle=0]{Fig12.eps}}
\mbox{\includegraphics*[width=3.in, angle=0]{Fig13.eps}}
\mbox{\includegraphics*[width=3.in, angle=0]{Fig14.eps}}
\caption{Results from the SPT, the present model and computer simulation
for HR fluid with $\kappa=3$. 
(a) equation of state resulting from SPT (dashed 
line), our proposal (solid line) and MC simulation results
(filled circles). The arrow indicates the packing fraction of
the I-N transition as estimated by simulation, while the filled square 
indicates the location of the I-N bifurcation point predicted by the present 
model; (b) behaviour of the uniaxial (solid line) and tetratic (dashed line) 
order parameters with packing fraction. Results from SPT and our model 
are indicated by the corresponding label. Symbols are simulation
results for the order parameters (open symbols: uniaxial, filled
symbols: tetratic); (c) orientational distribution functions at the 
coexistence packing fraction for the uniaxial nematic phase,
from SPT (dashed line) and the present model (solid line).}
\label{fig7}
\end{figure}

\begin{figure}
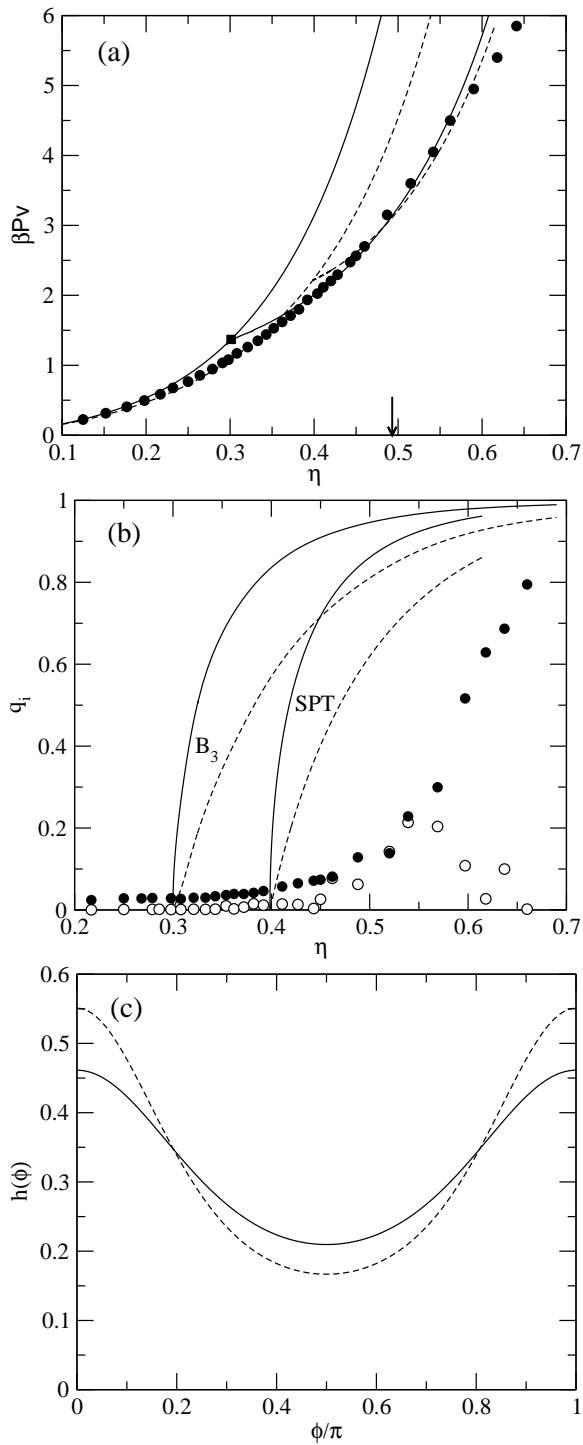

\mbox{\includegraphics*[width=3.in, angle=0]{Fig15.eps}}
\mbox{\includegraphics*[width=3.in, angle=0]{Fig16.eps}}
\mbox{\includegraphics*[width=3.in, angle=0]{Fig17.eps}}
\caption{Same as in Fig. \ref{fig7} but for 
$\kappa=9$. In (c) are shown the orientational distribution 
functions calculated at packing fractions separated from 
the bifurcation packing fraction a relative distance ($\Delta=0.01467$)
equal to that between the isotropic and nematic coexisting packing 
fractions for $\kappa=3$.}
\label{fig8}
\end{figure}

In order to appreciate more deeply the differences between the SPT and the
new theory, we now compare the EOS for the isotropic and orientationally ordered
phases and the behaviour of the order parameters, $q_1,q_2$, with packing fraction.
The latter are defined by 
\begin{eqnarray}
q_i=\int d\phi \cos{\left(2i\phi\right)} h(\phi),\hspace{0.7cm}i=1,2
\label{OPs}
\end{eqnarray}
with $q_1$ the uniaxial order parameter and $q_2$ the tetratic order parameter. 
The comparison is done in Figs. \ref{fig7}(a-c) for the case $\kappa=3$ and in 
Figs. \ref{fig8} (a-c) for $\kappa=9$. Also, the MC simulation 
results are shown. In the case of the EOS, both theories severely overestimate
the pressure in the nematic regime when $\kappa=3$; however, the
transition point, as mentioned previously, is much better reproduced 
when three-body correlations are included in the theory. For the longer 
particles the pressures are better reproduced, but the location and
nature of the transition from the isotropic to the nematic phase are
not correct; as already mentioned, if the uniaxial nematic phase is
not taken into account, three-body correlations seem to be very important
in promoting tetratic order in the isotropic phase. These correlations alone,
when higher-order correlations are not considered, probably overemphasise 
the relative stability of the uniaxial nematic phase with respect to the
tetratic phase in the case of long particles, and cause a premature 
instability of the latter as particles become longer. 

Comparison of
the orientational distribution functions in the case $\kappa=3$ indicates
again the role of three-body correlations. Fig. \ref{fig8} shows the
corresponding function for the uniaxial nematic phase that coexists with
the tetratic (new theory) or isotropic (SPT) phase; even though the new 
theory predicts a much lower transition density than SPT, tetratic ordering 
is much more pronounced in the new theory since at this value of $\kappa$ the 
tetratic phase is still stable. 

A point worth mentioning is the identification of the value of aspect ratio
where the tetratic phase is no longer stable. The nonequilibrium macroscopic
experiments  by Narayan et al. \cite{Narayan} find substantial tetratic
correlations in cylindrical particles with aspect ratio $\kappa=12.6$. Our present 
simulation data are not sufficiently detailed to give conclusive results. However, 
data for $\kappa=7$ (not shown) and $\kappa=9$ seem to be compatible with N$_t$ stability: 
the value of the uniaxial order parameter $q_1$ is compatible with zero 
in the whole density range studied. In the case $\kappa=9$ [Fig. \ref{fig8}(b)], 
however, there seems to be some tendency in the uniaxial order parameter
to increase from zero. 

Nevertheless, it is very difficult
within our present analysis, to settle this question. It is in fact 
difficult to distinguish between the N$_t$ and N$_u$ nematic phases, since
the latter exhibits substantial tetratic correlations even for the
longer particles considered ($\kappa=9$). 
Before further work is undertaken, all we can say conclusively from the 
simulation results is that at some packing fraction the isotropic
phase begins to display substantial tetratic order in a rather abrupt manner;
whether this order corresponds to uniaxial or strictly tetratic nematic phases
is a matter that would require more simulation work using e.g. larger systems.
Our limited study is only intended to provide approximate phase boundaries
for systems of particles with various aspect ratios (note that previous 
simulation 
work on this and related systems \cite{Frenkel,Frenkel1,Donev} were more 
detailed, but restricted to a particular aspect ratio).

\begin{figure}
\mbox{\includegraphics*[width=3.5in, angle=0]{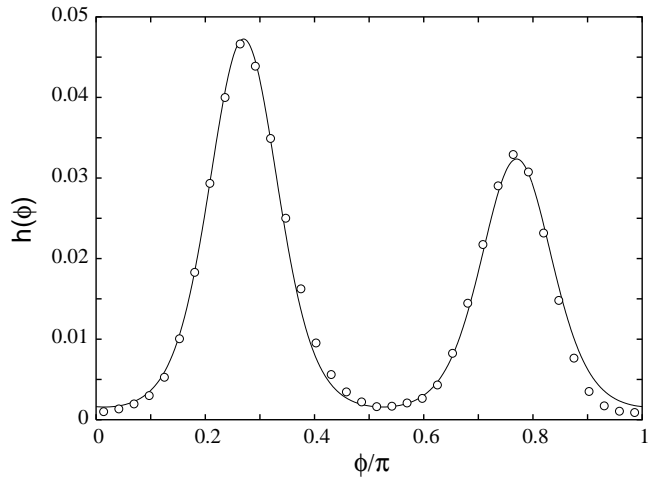}}
\caption{Orientational distribution function $h(\phi)$ from simulation for the
case $\kappa=9$ and packing fraction $\eta=0.618$. Symbols: simulation
data. Line: best fit to Eqn. (\ref{Varia}). Resulting values for the
order parameters are $q_1=0.153$ and $q_2=0.628$.}
\label{Ajuste}
\end{figure}

\subsection{Dependence on EOS adopted}

\begin{figure}
\mbox{\includegraphics*[width=3.4in, angle=0]{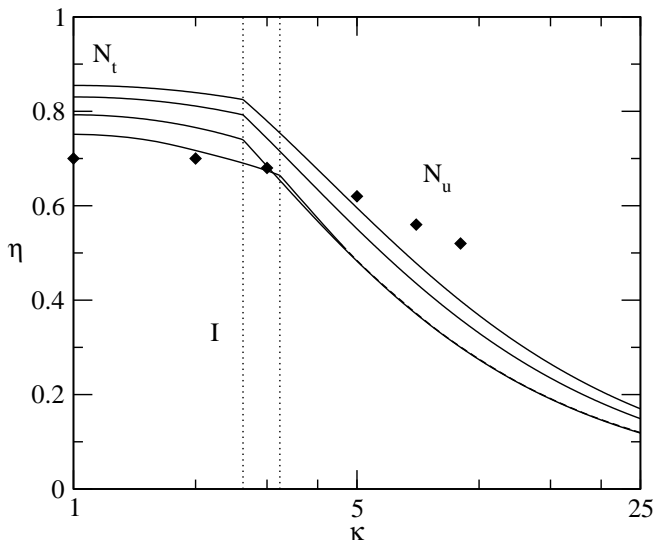}}
\caption{Spinodals of the transitions between the isotropic and orientational ordered phases. From top to 
bottom are shown the results from the SPT, the EOS with the approximated $b_3$ (the Boublik, and our proposal), and from the EOS with the exact $b_3$. The dotted lines represent the position of the tricritical points 
common to all the approximations and that corresponding to the exact $B_3$. Also are shown the simulation 
results.}
\label{fig6b}
\end{figure}

It should be noted that the packing fraction values of the I-N$_{u,t}$ phase transitions 
depend sensitively on the approximation used for the EOS of the HR fluid. This can be easily shown  
if we approximate the third virial coefficient 
as a function of the second virial 
coefficient using the relations 
\begin{eqnarray}
\gamma=b_2,\quad \beta \gamma^2=b_3-\alpha b_2,
\label{acta}
\end{eqnarray}
which can be easily obtained from (\ref{segundo}) and (\ref{tercero})
and are only valid for the isotropic fluid. From (\ref{acta}) 
we obtain   
\begin{eqnarray}
b_3=\alpha b_2+\beta b_2^2,
\label{approx}
\end{eqnarray} 
which can be used as an approximation of the third 
virial coefficient of the nematic fluid. 
Thus, inserting the above expression 
in Eq. (\ref{taken}), and carrying out the bifurcation analysis
described in Section IIC, we arrive at 
\begin{eqnarray}
\Delta\varphi_k\approx\frac{h_k^2}{4}\left[1-\Psi(y)b_2^{(k,k,0)}\right],\quad k=1,2 \label{reason}\\
\Psi(y)=y+\left[\alpha-2+2\beta\frac{(\kappa+1)^2}{\pi\kappa}\right]\theta(y),
\end{eqnarray}   
for the second order expansion of the free-energy difference between 
the I and N phases at the bifurcation point; here the subindex $k=1,2$ 
labels the N$_u$ and N$_t$ phases, respectively. Solving equation
$\Delta\varphi_k=0$ for $\eta(\kappa)$, we find the spinodals
shown in Fig. \ref{fig6b} for different values of $(\alpha,\beta)$
corresponding to those of SPT, Boublik's proposal, and our proposal for the 
EOS of the isotropic fluid. In the same figure the spinodal line resulting 
from the EOS with the exact third virial coefficient is also plotted. 
Comparing 
the latter with those obtained from the different approximations 
embodied in (\ref{approx}), we conclude that the location of the 
I-N$_u$ tricritical point changes only if one uses the exact three-body 
correlations. The reason for this behavior can be elucidated from Eqn. 
(\ref{reason}): the values of $(\eta^*,\kappa^*)$ calculated from 
$\Delta\varphi_1=\Delta\varphi_2=0$ gives us $\kappa^*=(3+\sqrt{5})/2$,
independent on the choice of $(\alpha,\beta)$ as the function $\Psi(y)$ 
does not depends on $k$.

\section{Conclusions}
The main results presented in this article can be summarized as follows.
(i) The inclusion of many- (higher-than-two) body correlations in 
two-dimensional systems of hard anisotropic bodies is of crucial importance 
in order to adequately describe the phase behavior of these systems. 
In two-dimensions two-body interactions are not enough to make quantitative 
predictions of their phase behavior, a crucial difference with respect to
three-dimensional systems. This conclusion is supported by simulation
results. ii) We have proposed an EOS and a corresponding
free energy density functional for fluids of hard rectangles that 
incorporates three-body correlations. While 
predicting pressures for the isotropic and nematic fluids which are too high 
when compared with simulation values, the theory gives values for the coexistence 
densities of the I-N$_t$ transition that 
compare fairly well with the simulation 
results for small values of $\kappa$.
A shortcoming of the theory is that the third virial coefficient,
which is incorporated exactly, has to be evaluated numerically beforehand.
This is a practical, not fundamental, limitation of the theory, which 
can be circumvented in all cases (i.e. for all different particle
geometries in two dimensions). 

A striking prediction of the theory is
the stability of a tetratic phase, in good quantitative agreement with
simulations for low particle aspect ratio. We conclude from this that
the four-fold correlations present in this phase are basically taken care 
of by our third-virial coefficient based theory, but not by the usual 
scaled-particle theory, which incorporates only two-body correlations.
More at variance with simulation results is the case of high aspect
ratios. In this regime the stability of the uniaxial nematic phase 
is overestimated with respect to the isotropic fluid. 

As a final comment, we must say that no attention has been paid
to non-uniform phases (smectic, columnar or solid) in the present work.
Previous studies by our group \cite{Martinez-Raton}, based on a 
density-functional theory which recovers SPT in the limit of spatially uniform
phases and combines Onsager and fundamental-measure theories,
indicate that the tetratic phase is preempted (in the sense of
bifurcation theory, i.e. spinodal lines) by a spatially ordered phase.
Inclusion of three-body correlations could severely affect
this result since these correlations may affect both phases differently.
In fact, simulations available so far support the conclusion that the
tetratic phase may be stabilised prior to crystallisation.
However, even in the case that it were possible to construct a density 
functional, suitable for such non-uniform phases, and incorporating three-body 
correlations, the effort involved in the minimization
with respect to the full density profile $\rho({\bf r},\phi)$ would 
be rather huge. Work along this avenue is now in progress in our group.

\begin{acknowledgments}
Y.M.-R. was supported by a Ram\'on y Cajal research contract from the
Ministerio de Ciencia y Tecnolog\'{\i}a (Spain).
This work is part of the research
Projects No. BFM2003-0180, FIS2005-05243-C02-01, FIS2005-05243-C02-02 and
FIS2004-05035-C03-02 of the Ministerio de Educaci\'on y Ciencia
(Spain), and S-0505/ESP-0299 of Comunidad Aut\'onoma de Madrid.
\end{acknowledgments}

\section{Appendix}

\subsection{Bifurcation analysis}
In this section we introduce the formalism
that allowed us to calculate the 
spinodal instabilities and the order of the phase transitions
between the isotropic and orientational ordered phases. This formalism 
is quite general, in the sense that it is independent of the geometry of 
particles, and includes two- and three-particle correlations. As usual, 
the analysis starts from an order-parameter expansion of the free-energy 
difference ($\Delta\varphi$) between the bifurcated (orientationally ordered) 
and the parent (isotropic) phases about the bifurcation point, and then the 
evaluation of the inverse isothermal compressibility ($\varkappa^{-1}$) 
of the bifurcated phase at the same point. The existence of a tricritical point,
at which the order of the transition changes from second to first order,  
is predicted from the first change of sign (from positive to negative) 
of $\Delta\varphi$ or $\varkappa^{-1}$. The starting point of the bifurcation 
analysis is to assume that the orientational distribution function near the  
I-N bifurcation point can be approximated as a Fourier series in small 
amplitudes $h_k\sim \epsilon^k$ (where $\epsilon$ is an small parameter), 
truncated at second order, i.e. 
\begin{eqnarray}
h(\phi)\approx \frac{1}{\pi}\left(1+h_1\cos 2\phi +h_2\cos 4\phi\right).
\end{eqnarray}
Inserting this expression into (\ref{taken}) and (\ref{id}), we obtain the 
difference between the nematic and isotropic free energies per particle as
\begin{eqnarray}
&&\Delta \varphi\equiv \varphi_{\rm{N}}-\varphi_{\rm{I}}
\approx Ah_1^2+Ch_1^2h_2+Dh_2^2+Eh_1^4,
\label{ener}
\end{eqnarray} 
where the coefficients $A,C,D,E$ have the form
\begin{eqnarray}
A&=&\frac{1}{4}\left[1-yb_2^{(1,1)}-\theta(y)\left(b_3^{(1,1,0)}
-2b_2^{(1,1)}\right)\right],\label{laa}\\
C&=&-\frac{1}{8}\left[1+2\theta(y)b_3^{(1,1,2)}\right],\label{lac}\\
D&=&\frac{1}{4}\left[1-yb_2^{(2,2)}-\theta(y)\left(b_3^{(2,2,0)}
-2b_2^{(2,2)}\right)\right],\label{lad}\\
E&=&\frac{1}{32}\label{lae},
\end{eqnarray}
and we have defined $y=\eta/(1-\eta)$. $\theta(y)=y-\ln(1+y)$ is the same 
function as in (\ref{exceso}), but here in terms of the new variable $y$. 
The coefficients 
\begin{eqnarray}
b_i^{(k_1,\dots,k_i)}&=&
-\frac{4}{\pi^iv^{i-1}}\left[\prod_{l=1}^i\int_0^{\pi}d\phi_l\cos 2k_l\phi_l
\right]\nonumber \\
&\times&{\cal K}(\phi_1,\dots,\phi_i),\quad i=2,3,\quad k_j\in \mathbb{N},  
\end{eqnarray}
have also been defined, which originate from two- ($i=2$) and three- ($i=3$) 
body correlations. For $i=2$ one can obtain analytic results which, 
for the specific case of HR, give
\begin{eqnarray}
b_2^{(k,k)}=\frac{2}{(4k^2-1)\pi}\frac{\left(L+(-1)^k\sigma\right)^2}{v}
\label{anal}
\end{eqnarray} 
If we set $\theta(y)=0$ in (\ref{laa})-(\ref{lad}) the SPT result is recovered.
Minimizing the free energy difference (\ref{ener}) with respect to $h_2$, we
obtain $h_2$ as a function of $h_1$ [$h_2=-Ch_1^2/(2D)$] which,
inserted in (\ref{ener}), results in 
\begin{eqnarray}
\Delta\varphi&=&Ah_1^2+Bh_1^4,\label{efect}\\ 
B&=&E-\frac{C^2}{4D}\label{otrab}
\end{eqnarray}
where $A$ and $B$ are functions of the variable $y$.
Minimizing Eq. (\ref{efect}) with respect to $h_1$, and taking into account 
the expansion of $y$ about its bifurcation value $y^*$, i.e. 
$y\approx y^*+y^{(2)}h_1^2$, we arrive at 
\begin{eqnarray}
\frac{\partial \Delta\varphi}{\partial h_1}=2h_1\left[
A^*+\left(2B^*+A_y^*y^{(2)}\right)h_1^2\right]=0,
\label{solv}
\end{eqnarray} 
where $A_y$ is the first derivative of $A$ with respect to $y$, and 
the asterisk on $A,B$, and $A_y$ means that these functions are evaluated at 
the bifurcation point $y^*$. Solving (\ref{solv}) order by order, we 
obtain two equations: 
\begin{eqnarray}
A^*&=&0,\label{adin}\\
y^{(2)}&=&-\frac{2B^*}{A_y^*},
\label{dva}
\end{eqnarray}
the first one allowing to find $y^*$, and hence the packing 
fraction $\eta^*(\kappa)$, as a function of the particle aspect ratio 
$\kappa=L/\sigma$, i.e the spinodal line of the I-N phase transition. 
Expanding (\ref{efect}) about the bifurcation point, and using 
(\ref{adin}) and (\ref{dva}), we obtain 
\begin{eqnarray}
\Delta\varphi=-B^*h_1^4,
\label{final}
\end{eqnarray}
which indicates that the I-N transition is of first order if $B^*<0$. 
Eqn. (\ref{final}) can be written in a different, more convenient form,
with use of $h_1^2=(y-y^*)/y^{(2)}$ and Eqn. (\ref{dva}), which results in 
\begin{eqnarray}
\varphi_{\rm{N}}=\varphi_{\rm{I}}^*-\frac{\left(A_y^*\right)^2}{4B^*}
(y-y^*)^2.
\label{otro}
\end{eqnarray}  
Using the definition of the inverse isothermal compressibility, 
$\varkappa^{-1}=\rho \partial(\beta P)/\partial\rho$, 
in terms of the $y$ variable,
\begin{eqnarray}
\varkappa^{-1} v=y(1+y)\frac{\partial}{\partial y}\left(
y^2\frac{\partial \varphi}{\partial y}\right),
\end{eqnarray}
together with Eqn. (\ref{otro}), we find 
\begin{eqnarray}
\left(\varkappa^{-1}_{\rm{N}}v\right)^*=
\left(\varkappa^{-1}_{\rm{I}}v\right)^*-\left(y^*\right)^3(1+y^*)
\frac{\left(A_y^*\right)^2}{2B^*},
\end{eqnarray}
where  
\begin{eqnarray}
\left(\varkappa_{\rm{I}}^{-1} v\right)^*&=&
y^*\left(1+y^*\right)\left(1+2y^*b_2\right)\nonumber\\&+&
\frac{\left(y^*\right)^3\left(3+2y^*\right)}{1+y^*}(b_3-2b_2).
\end{eqnarray}
The existence of a tricritical point, at which the I-N transition changes 
from second to first order as particles change from large to small 
aspect ratios, can be found for a value of the aspect ratio $\kappa^*$
satifying $\kappa^*=
\rm{max}\left(\kappa_1,\kappa_2\right)$, where $\kappa_j$ ($j=1,2$) are 
the solutions to the equations
\begin{eqnarray}
B^*(\kappa_1)=0,\quad \left(\varkappa_{\rm{N}}^{-1}v\right)^*(\kappa_2)=0
\label{dosenuna}
\end{eqnarray}
The preceding analysis with $h_k\neq 0$ ($k=1,2$) corresponds 
to the bifurcation 
analysis of the transition between the isotropic and the uniaxial nematic 
phase N$_u$. If $h_1=0, h_2\neq 0$, the bifurcating phase is a tetratic nematic 
phase N$_t$. To carry out the bifurcation analysis for the I-N$_t$ transition, 
we can use exactly the same formalism, except that
we have to make the substitutions $h_k\to h_{2k}$, 
and $b_i^{(k_1,\dots,k_i)}\to b_i^{(2k_1,\dots,2k_i)}$ in Eqns. (\ref{ener})-
(\ref{lad}).

Taking the Onsager limit $\kappa\to \infty$ in Eqn. (\ref{laa}) and considering 
that, in the asymptotic limit [see Eqn. (\ref{undos})], the coefficients 
$b_2^{(1,1)}$ and $b_3^{(1,1,0)}$ are of order $\kappa$ and $\kappa^2$, 
respectively, the condition (\ref{adin}) is equivalent to solving a 
second-order equation with respect to the reduced density 
$\rho_r\equiv \rho^* L^2$, with the solution
\begin{eqnarray}
\rho_r=\frac{1}{\tau^*}\left(\sqrt{1+3\pi\tau^*}-1\right),
\label{rho_r}
\end{eqnarray}
where we have defined the coefficient
\begin{eqnarray}
\tau^*=\lim_{\kappa\to\infty} \tau(\kappa),\quad 
\tau(\kappa)=\frac{3\pi}{2}\frac{b_3^{(1,1,0)}(\kappa)}{\kappa^2}.
\label{tau}
\end{eqnarray} 
The limit $\tau^*\to 0$ of Eq. (\ref{rho_r}) recovers the SPT result 
$\rho_r=3\pi/2$. Also $\rho_r(\tau^*)$ as a function of $\tau^*$ 
is a monotonically decreasing 
function whose domain and image are $[-1/3\pi,\infty)$ and 
$(0,3\pi]$, respectively. Thus if $\tau^*>0$ ($\tau^*<0$), 
the I-N transition in a 
two-dimensional hard-needle fluid occurs at a reduced density in the
interval $(0,3\pi/2]$ ($[3\pi/2,3\pi)$).   

\subsection{Calculation of $B_3$}

The third virial coefficient $B_3(\{\lambda_{\tau}\})$ was obtained 
by MC integration using, for the orientational distribution function,
the form  
\begin{eqnarray}
h(\phi)=C \exp{\left(\lambda_1\cos{2\phi}+\lambda_2\cos{4\phi}\right)}
\label{Varia}
\end{eqnarray}
which contains two free parameters, $\lambda_1$ and $\lambda_2$. 
The MC data for $B_3$ were obtained for each value of $\kappa$ explored
and for fixed values of $\lambda_1$ and $\lambda_2$. The technique followed 
was a generalization of the standard method for
isotropic fluids \cite{Hoover}: each step involved generating angles for the 
three rectangles [see Eqn. (\ref{kernel3})] and positions for two of them 
[the first rectangle is placed at the origin, see Eqn. (\ref{kernel3})]. 
Since the $B_3$ coefficient involves a single irreducible cluster integral
where all three rectangles overlap, the positions of the second and third
rectangles were chosen within their respective excluded volumes with the
first rectangle to insure overlap, and only one overlap condition 
(second with third rectangles) had to be checked. Angles were
generated using an acceptance-rejection method according to the
angular distribution function $h(\phi)$ corresponding to the values of 
$\lambda_1$ and $\lambda_2$ (the method was checked by computing the
second virial coefficient $B_2$ for the isotropic case, which can be
compared with the exact result, and also the third virial coefficient
for some special particle orientations where this coefficient is analytic;
in this case the computed values of $B_2$ and $B_3$ for high values of
$\lambda_1$ and $\lambda_2=0$ tended to the correct value). A single
MC step involves generating one chain of rectangles, and  
$1-2\times 10^{7}$ MC steps were used in the calculations for each
set of values of $\lambda_1,\lambda_2$.

We generated numerical values of $B_3$ at a collection 
of mesh points on a rectangular region of the $\lambda_1-\lambda_2$ plane. 
The extension of this region was chosen according to the values of the packing 
fraction. In any case it contained the origin ($\lambda_1=\lambda_2=0$) to allow
for the isotropic phase. In some cases the region $[-1,1]\times[-1,1]$
did suffice; in others, higher minimum and maximum values for the parameters
were needed, especially when a first-order transition was detected.
The mesh interval was typically $\Delta\lambda=0.1$, with finer meshes
when required. In order to use these data in a practical way, the data
were fitted in two different ways. One involves constructing a polynomial 
$P_{N}(\lambda_1,\lambda_2)=\sum_{n=0}^N
\sum_{m=0}^nc_{nm}\lambda_1^m\lambda_2^{n-m}$ by a least-square procedure. 
Symmetry considerations require some terms of this polynomial expansion 
not to appear, and the corresponding coefficients $c_{nm}$ were
taken to be zero. The degree of the polynomial was typically 
in the range $8-10$. In the case of the I-N$_t$ transition, which only
involves $q_2$ (and hence $\lambda_2$), calculations were also done using
fits to a polynomial depending only on $\lambda_2$ (since necessarily
$\lambda_1=0$). Results are consistent with the previous results based
on a full fitting.

For high packing fractions a fit to a function in the order parameters
$(q_1,q_2)$ is more suitable since their values are close to one, whereas
the $\lambda$ parameters grow without limit. However, the dependence of $B_3$
on $(q_1,q_2)$ is strong. We found it useful to use a combination of
polynomials in the $q$'s and factors of the form $(q_i\pm 1)^n$, with
$n$ a power whose value is optimized in the fit.

\subsection{Minimization of the free-energy functional}

The minimizations were done using a variational scheme. 
An important question is how accurate is the variational function (\ref{Varia}).
We can assess the quality of this function by comparing with 
simulation results for the distribution function $h(\phi)$. Fig. \ref{Ajuste}
shows a distribution-function histogram obtained from a constant-pressure
MC simulation, over $2\times 10^6$ steps, of a fluid with $\kappa=5$ at 
pressure $P\sigma^2/kT=1.4$.
This is clearly a nematic phase with tetratic order (whether this 
corresponds to a uniaxial or purely tetratic phase is a different matter;
extremely long runs are probably needed to fully equilibrate the system. 
For the present purpose this is of no importance). A least-square
fitting to the variational function gives $\lambda_1=0.033$, 
$\lambda_2=1.217$ ($q_1=0.008$, $q_2=0.523$), which results in the 
function represented in the
figure. Histograms exhibiting more structured orientational 
order can be similarly fitted with comparable accuracy. The function 
(\ref{Varia}) is therefore suitable as a variational function.

The nematic order parameters can be related to the variational 
parameters via Eqn. (\ref{OPs}).
There is a one-to-one correspondence between the sets
$(q_1,q_2)$ and $(\lambda_1,\lambda_2)$.
In our calculations the function 
$\varphi(q_1,q_2;\kappa,\eta)$ was then minimized with respect to
$q_1,q_2$, using a standard Newton-Raphson technique.
All transitions are obtained as second-order transitions, except
in the approximate interval $3\alt\kappa\alt 5$ where discontinuous
transitions were found. Of course these results are consistent with
those from bifurcation theory, which is otherwise better suited for
the calculation of the tricritical points since it is not tied to
any variational scheme. The value of the functional minimization 
can be better appreciated in the case of the N$_t$-N$_u$ transition,
which cannot easily be obtained using bifurcation theory.

The results of the minimisation using the resulting
fitted function for $B_3$ are very sensitive to 
details such as mesh interval and degree of fitting polynomial. A detailed study
of the whole procedure, including fine tuning of the above parameters,
was therefore necessary. The accuracy of the results is sufficient to
locate the phase transitions with respect to packing fraction, and to
discriminate between first- and second-order phase transitions when
the system is far from the tricritical points, but the exact density
gap in first-order transitions (which is otherwise small) could not
be obtained with the present numerical implementation. 

\subsection{Some details on MC simulations}

Our constant-pressure MC simulations were performed on systems of 
$\sim 10^3-10^4$ HR particles, using rectangular cells and periodic
boundary conditions. The equation of state, orientational 
distribution function and nematic order parameters were obtained during the
course of these simulations. The simulations were run typically over 
$\sim 10^7$ MC steps for equilibration and $\sim 2\times 10^7$
MC steps for averaging (slow orientational dynamics, especially in the
case of long particles, require longer runs than in the isotropic phase).
The value of the pressure is fixed at some constant value. The average
density (or packing fraction $\eta$) is obtained, which gives the equation of
state $P(\eta)$. 

The orientational order is obtained from the eigenvalues and eigenvectors of
the order matrix, defined for a given particle configuration as
\begin{eqnarray}
S_{ij}=\frac{1}{N}\sum_{k=1}^N \left(2\hat{n}_i^k\hat{n}_j^k-\delta_{ij}\right)
\end{eqnarray}
where $\hat{\bf n}^k$ is a unit vector along the long axis of the $k$th 
particle. These is to be averaged over MC configurations.
The eigenvector associated with the largest eigenvalue gives the
direction of the primary director, and this eigenvalue is the uniaxial
order parameter, $q_1$, which can also be calculated from the average
\begin{eqnarray}
q_1=\frac{1}{N}\sum_{k=1}^N \left<\cos{2\left(\phi_k-\phi_0\right)}\right>
\end{eqnarray}
where $\phi_0$ is the polar angle of the director (which depends on the
configuration) and $\phi_k$ the polar angle of the particle long axis, both
with respect to some fixed direction in the plane. 
The tetratic order parameter can now be calculated from
a similar equation, with the factor 2 in the cosine substituted by 4.
Also, the orientational distribution function $h(\phi)$
was calculated as a histogram, using the angle $\phi_0$ as the origin.
From this the order parameters $q_i$ can also be obtained from Eqns. 
(\ref{OPs}). Yet another route is provided by the asymptotic value of 
the orientational correlation functions. No attempt was made at
calculating these functions in the present work.

The function $h(\phi)$ is always seen to have two maxima: a 
primary and a secondary maximum, separated by $\pi/2$; in the uniaxial
nematic phase these maxima should have different heights, whereas in
the tetratic phase their heights should be statistically equal. 
Due to many effects that
affect the simulations, distinguishing these two
situations is a rather delicate problem and our limited study did not 
allow identification of the true nature of the nematic phase. 
Questions such as effect of
boundary conditions, system size, etc, may be of paramount importance
in this analysis. For example, the rectangular periodic boundary conditions used
in this work could be artificially promoting tetratic ordering in the system.

As is characteristic of two-dimensional systems with continuous symmetries,
the orientationally ordered phases of the present model seem to
exhibit quasi-long-range order at long distances \cite{Donev}. This means, in 
particular, that the order parameters may show a strong system-size
dependence. This point has not been considered at all, since our only
aim was to establish approximate phase-stability boundaries that could
serve as a test bed against which the (otherwise approximate) theories
could be tested.

Finally, since this was not the aim of this work, and also
due to the difficulties involved in dealing with possibly multiply
degenerate structures, both periodic and nonperiodic, crystalline 
configurations have not been studied in any detail; however, freezing into 
glassy states in compression runs were observed to occur (these states were 
characterised by extremely low particle diffusion) at high density. These 
densities at quite close to those at which melting into a nematic phase from 
crystalline configurations are observed to occur in expansion runs
(starting from crystals with various types of
packing --particles perfectly aligned on a rectangular lattice, 
square clusters on square lattices, various random tilings, etc.)
A full discussion of this issue can be found in Ref. \cite{Donev}.


\begin{references}
\bibitem{Schaklen}H. Schlacken, H. -J. Mogel, and P. Schiller, 
Mol. Phys. {\bf 93}, 777 (1998).
\bibitem{Martinez-Raton}Y. Mart\'{\i}nez-Rat\'on, E. Velasco, and 
L. Mederos, J. Chem. Phys.  {\bf 122}, 064903 (2005).
\bibitem{Frenkel1}K. W. Wojciechowski and D. Frenkel, Comp. Meth. Sci. Tech. 
{\bf 10}, 235 (2004).
\bibitem{Donev}A. Donev, J. Burton, F. H. Stillinger, and
S. Torquato, Phys. Rev. B {\bf 73}, 054109 (2006).
\bibitem{PTCDA}M. M\"obus, N. Karl and T. Kobayashi, J. Crys. Growth {\bf 116},
495 (1992); L. Nony, R. Bennewitz, O. Pfeiffer, E. Gnecco, A. Baratoff, 
E. Meyer, T. Eguchi, A. Gourdon and C. Joachin, Nanotechnology {\bf 15}, 
S91-S96 (2004); H. Proehl, T. Dienel, R. Nitsche and T. Fritz, Phys. Rev. 
Lett. {\bf 93}, 097403 (2004). 
\bibitem{Narayan} V. Narayan, N. Menon, and S. Ramaswamy, 
J. Stat. Mech. P01005 (2006)
\bibitem{Onsager} L. Onsager, Ann. N. Y. Acad. Sci. {\bf 51}, 627 (1949).
\bibitem{Tarjus} G. Tarjus, P. Viot, S. M.Ricci, and J. Talbot, 
Mol. Phys. {\bf 73}, 773 (1991).
\bibitem{Reiss} H. Reiss, H. L. Frisch, and J. L. Lebowitz, J. Chem. Phys. 
{\bf 31}, 369 (1959).
\bibitem{Cotter} M. A. Cotter and D. E. Martire, J. Chem. Phys. 
{\bf 52}, 1902 (1970); J. Chem. Phys. {\bf 53}, 4500 (1970); M. A. 
Cotter and D. C. Wacker, Phys. Rev. A {\bf 18}, 2669 (1978). 
\bibitem{Lasher} G. Lasher, J. Chem. Phys. {\bf 53}, 4141 (1970).
\bibitem{Barboy} B. Barboy and W. Gelbart, J. Chem. Phys. {\bf 71}, 
3053 (1979).
\bibitem{Isihara} A. Isihara, J. Chem. Phys. {\bf 18}, 1446 (1950).
\bibitem{Kihara} T. Kihara, Rev. Mod. Phys. {\bf 25}, 831 (1953).
\bibitem{Boublik} T. Boublik and I. Nezbeda, Coll. Czech. Chem. 
Comm. {\bf 51}, 2301 (1986).
\bibitem{Boublik1}T. Boublik, Molec. Phys. {\bf 63}, 685 (1988).
\bibitem{Rigby} M. Rigby, Mol. Phys. {\bf 78}, 21 (1993).
\bibitem{Frenkel} D. Frenkel and R. Eppenga, Phys. Rev. A {\bf 31}, 1776 (1985).
\bibitem{Hoover}F. H. Ree and W. G. J. Hoover, J. Chem Phys. {\bf 40}, 939 (1964); D. Frenkel, J. Phys. Chem. {\bf 91}, 4912 (1987).
%\bibitem{hard_disks} E. Helfand, H.L. Frisch, and J.L. Lebowitz, J. Chem. Phys. {\bf 34}, 1037 (1961).
%\bibitem{Dijkstra1} M. Dijkstra, R. van Roij, and R. Evans, Phys. Rev. Lett.
%{\bf 81}, 2268 (1998); Phys. Rev. E {\bf 59}, 5744 (1998).
%\bibitem{Almarza} N. G. Almarza and E. Enciso, 
%Phys. Rev. E {\bf 59}, 4426 (1998).
%\bibitem{Buhot} A. Buhot and W. Krauth, Phys. Rev. Lett. {\bf 80}, 3787 
%(1997).
%\bibitem{Shuri} Y. Mart\'{\i}nez-Rat\'on, J. A. Cuesta, 
%Phys. Rev. E {\bf 58}, R4080 (1998).
%\bibitem{Mederos} E. Velasco, G. Navascu\'es and L. Mederos, 
%Phys. Rev. E {\bf 60}, 3158 (1999).
%\bibitem{Lafuente} L. Lafuente and J. A. Cuesta, Phys. Rev. Lett. 
%{\bf 89}, 145701 (2002).
%\bibitem{Biben} T. Biben, P. Blandon and D. Frenkel, J. Phys.: Condens. Matter 
%{\bf 8}, 10799 (1996).
%%%%% mezcla de varillas y discos%%%%%%%%
%\bibitem{Roij1}R. van Roij and B. Mulder, J. Phys. II France {\bf 4}, 
%1763 (1994).
%\bibitem{Wensink} H. H. Wensink, G. J. Vroege, and H. N. W. 
%Lekkerkerker, J. Chem. Phys. {\bf 115}, 7319 (2001).
%\bibitem{Perera} S. Dubois and A. Perera, J. Chem. Phys. {\bf 116}, 6354 
%(2002); A. Perera, K. Cassou, F. Ple and S. Dubois, 
%Mol. Phys. {\bf 100}, 3409 (2002); A. Perera, J. Mol. Liq. {\bf 109}, 
%73 (2004).
%\bibitem{Varga1} S. Varga, A. Galindo, and G. Jackson, J. Chem. Phys. 
%{\bf 117}, 7207 (2002); A. Galindo, A. J.Haslam, S. Varga, G. Jackson, 
%A. G. Vanakaras, D. J. Photinos, and D. A. Dunmur, J. Chem. Phys. 
%{\bf 119}, 5216 (2003).
%\bibitem{Martinez-Raton1}Y. Mart\'{\i}nez-Rat\'on and J. A. Cuesta, 
%J. Chem. Phys. {\bf 118}, 10164 (2003).
%%%%%%% mezcla de esferas y varillas %%%%%%%
%\bibitem{Schmidt}M.Schmidt and A. R. Denton, Phys. Rev. E {\bf 65}, 
%021508 (2002).
%%%%%%% mezcla de esferas y discos %%%%%%%%%
%\bibitem{Oversteegen} S. M. Oversteegen and H. N. W. Lekkerkerker, 
%J.Chem. Phys. {\bf 120}, 2470 (2003).
%%%%%%%% mezcla de varillas %%%%%%%
%\bibitem{Roij2}R. van Roij, and B. Mulder, Phys. Rev. E {\bf 54}, 
%6430 (1996); R. van Roij, B. Mulder and M. Dijkstra, Physica A {\bf 261}, 
%374 (1998).
%\bibitem{Dijkstra2}M. Dijkstra and R. van Roij, Phys. Rev. E {\bf 56},
%5594 (1997).
%\bibitem{Varga2} S. Varga, A. Galindo and G. Jackson, Mol. Phys. {\bf 101}, 
%817 (2003).
%%%%%% SPT for hard convex bodies %%%%%%%%
%\bibitem{Talbot} J. Talbot, J. Chem. Phys. {\bf 106}, 4696 (1997).
%\bibitem{Rosenfeld}Y. Rosenfeld, Phys.Rev. Lett. {\bf 63}, 980 (1989). 
%\bibitem{Tarazona} P. Tarazona, Phys. Rev. Lett. {\bf 84}, 694 (2000).
%\bibitem{Tenne} R. Tenne and E. Bergmann, Phys. Rev. A {\bf 17}, 
%2036 (1978); R. J. Bearman and R. M. Mazo, J. Chem. Phys. {\bf 88}, 1235 
%(1988); J. Chem. Phys. {\bf 91}, 1227 (1989); R. Mazo and R. J. Bearman, 
%J. Chem. Phys. {\bf 93}, 6694 (1990). 
%\bibitem{Mountain} R. Mountain and A. H. Harvey, J. Chem. Phys. {\bf 94}, 
%2238 (1991); F. Saija and P.V. Giaquinta, J. Chem. Phys. {\bf 117}, 
%5780 (2002).
%\bibitem{Raveche}R. F. Kayser and H. Ravech\'e, Phys. Rev. A {\bf 17}, 
%2067 (1978).
%\bibitem{Schoot} P. van der Schoot, J. Chem. Phys. {\bf 106}, 2355 (1997).
%\bibitem{Frenkel} M. A. Bates and D. Frenkel,
%J. Chem. Phys. {\bf 112}, 10034 (2000).
%\bibitem{Martinez-Raton2} Y.Mart\'{\i}nez-Rat\'on, E.Velasco, and L. 
%Mederos, J. Chem. Phys. {\bf 122}, 064903 (2005).
%\bibitem{Sluckin} A. Poniewierski and T. J. Sluckin, Phys. Rev. A {\bf 43}, 
%6837 (1991).
%\bibitem{kappa}Note that in the one-component limit the 
%condition $H^*=0$ should be substituted by 
% $\left(\varkappa_T^{-1}\right)^*=0$ 
%(where $\varkappa_T^{-1}=-V
%\displaystyle{\left(\frac{\partial P}{\partial V}\right)_T}$ is    
%the inverse isothermal compressibility).
%
\end{references}
\end{document}